\documentclass[aps,pra,superscriptaddress,10pt,twocolumn
]{revtex4-2}

\usepackage{ascmac,bm}
\usepackage{graphicx}
\usepackage{amsmath,amssymb,mathrsfs,titlesec,physics,amsfonts,here,enumerate,mathtools}
\usepackage{natbib}
\usepackage{url,textcase}
\usepackage{color}
\usepackage[svgnames]{xcolor}
\usepackage[colorlinks=true,citecolor=Green,linkcolor=Blue,urlcolor=Purple,linktocpage,unicode]{hyperref}

\newcommand{\e}{\mathrm{e}}
\renewcommand{\i}{\mathrm{i}}
\renewcommand{\d}{{\rm d}}

\newcommand{\R}{\mathbb{R}}
\newcommand{\N}{\mathbb{N}}

\begin{document}
\title{Redesigning the linear--quadratic--Gaussian cost function for feedback cooling of a quantum harmonic oscillator}
\author{Shuma Sugiura}
    \email{sugiura@cat.phys.s.u-tokyo.ac.jp}
    \affiliation{%
        Department of Physics, University of Tokyo, 7-3-1 Hongo, Bunkyo-ku, Tokyo, 113-8654, Japan}
    \affiliation{%
        Department of Physics, King's College London, Strand, London WC2R 2LS, United Kingdom}
\author{Koki Shiraishi}
    \affiliation{%
        Department of Physics, University of Tokyo, 7-3-1 Hongo, Bunkyo-ku, Tokyo, 113-8654, Japan}
\author{Masahito Ueda}%
    \affiliation{%
        Department of Physics, University of Tokyo, 7-3-1 Hongo, Bunkyo-ku, Tokyo, 113-8654, Japan}
    \affiliation{%
        Institute for Physics of Intelligence, University of Tokyo, 7-3-1 Hongo, Bunkyo-ku, Tokyo, 113-0033, Japan}
    \affiliation{%
        Fundamental Quantum Science Program (FQSP), TRIP Headquarters, RIKEN, Wako, 351-0198, Japan}
\date{\today}

\begin{abstract}
    Linear--quadratic--Gaussian (LQG) control is optimal only with respect 
    to a prescribed cost function, 
    the choice of which dictates the physical objective of the control. 
    We consider feedback cooling of a continuously monitored quantum harmonic oscillator 
    by shifting the minimum of its trapping potential. 
    In this setting, 
    the physically relevant cooling objective can be defined as minimizing the oscillator's energy 
    relative to the feedback-shifted potential. 
    In contrast, conventional LQG control evaluates the energy from a fixed origin 
    and thus fails to directly optimize this quantity. 
    To address this problem, we introduce a redesigned cost function 
    that explicitly accounts for the feedback-induced shift of the potential. 
    We then derive the corresponding optimal feedback law 
    and obtain an analytic expression for the minimum achievable steady-state phonon occupation number. 
    The redesigned LQG control achieves a lower occupation number 
    than low-pass-filter (LPF) feedback formulated for the same cooling objective. 
    While this improvement is minor at detection efficiencies currently attainable in experiments---%
    indicating that LPF feedback already delivers near-optimal cooling performance---%
    the advantage becomes pronounced as the detection efficiency approaches unity.
    In this regime, the redesigned LQG control provides an increasing advantage 
    for reaching the motional ground state at a finite measurement strength. 
    We clarify that the conventional and redesigned cost functions 
    represent distinct control objectives 
    rather than different implementations of the same optimization problem.
\end{abstract}

\maketitle

\section{Introduction} \label{sec:Intro}

Linear--quadratic--Gaussian (LQG) control 
provides an optimal feedback-control framework 
for linear systems subject to Gaussian 
noise~\cite{10.5555/531315,stengel1994optimal,lewis2012optimal,levine2018control}, 
with its optimality defined relative to a prescribed quadratic cost function. 
For instance, 
when the cost function represents the energy of a particle, 
minimizing thi cost corresponds directly to cooling the particle. 
LQG control has been widely employed in various high-precision engineering systems, 
including safety-critical aerospace 
platforms~\cite{gupta1980frequency,athans1986linear,garg1989turbofan,garg1991integrated,grewal1996robust,CHRIF2014245}, 
high-bandwidth industrial servo 
systems~\cite{hanselmann1988lqg,weerasooriya1995discrete,hu1999discrete,chen2002hard,panigrahi2015robust}, 
and adaptive optics for precision 
measurement~\cite{petit2008first,looze2008linear,petit2009linear,correia2010optimal}.

While LQG control was originally developed for classical systems, 
the formulation of a quantum analogue of the Kalman filter 
has enabled its extension to continuously monitored quantum systems~\cite{doherty1999feedback,wiseman2009quantum,ZHANG20171}. 
One notable application is measurement-based feedback cooling of mechanical motion, 
which has been experimentally demonstrated for a levitated nanoparticle 
using real-time state estimation~\cite{magrini2021real}. 
More broadly, 
ground-state cooling of levitated nanoparticles has been achieved 
via several cooling schemes~\cite{delic2020cooling,magrini2021real,kamba2022optical,piotrowski2023simultaneous}. 
As reviewed in Refs.~\cite{millen2020optomechanics,gonzalez2021levitodynamics}, 
cooling a levitated nanoparticle toward its motional ground state 
can enhance the sensitivity of sensors 
such as accelerometers~\cite{yang2024photon,kamba2026levitated}, 
weak-force detectors~\cite{hempston2017force,PhysRevLett.121.063602,liang2023yoctonewton}, 
and gravitational-wave detectors~\cite{PhysRevLett.110.071105,PhysRevLett.128.111101}. 
Ground-state cooling also serves as a crucial starting point for preparing nonclassical motional states, 
including macroscopic superposition states and motionally entangled states~\cite{gonzalez2021levitodynamics}. 
Recent demonstrations of quantum delocalization~\cite{rossi2025quantum} 
and squeezing~\cite{kamba2025quantum} 
further highlight the prospects of such quantum-state control.

Here, 
we model a levitated nanoparticle as a quantum harmonic oscillator 
and, for simplicity, restrict its motion to one dimension. 
We consider feedback cooling implemented by shifting the minimum of the trapping potential. 
Because this feedback operation alters the potential experienced by the oscillator, 
it changes the instantaneous ground state---the very state against which cooling should be assessed. 
For feedback schemes based on a low-pass filter (LPF), 
this feedback-induced modification of the potential has been incorporated 
into the effective energy used to quantify cooling~\cite{PhysRevE.111.014152,sugiura2025quantum}. 
We refer to such schemes as the LPF feedback.

The corresponding physical objective is therefore 
to minimize the oscillator energy with respect to the total, feedback-shifted potential, 
thereby bringing the oscillator closer to its instantaneous ground state. 
To our knowledge, 
LQG control formulated explicitly for this cooling objective 
remains unexplored. 
Previous applications of LQG control to continuously monitored harmonic motion 
have instead employed a cost function defined relative to a fixed reference point, 
alongside a separate penalty on the feedback operation~\cite{doherty1999feedback,magrini2021real}. 
Consequently, these two formulations target fundamentally distinct physical control objectives.

In this paper, 
we formulate the LQG control problem for feedback cooling 
with respect to the feedback-shifted potential. 
To accomplish this physical objective within the LQG framework, 
we redesign the cost function to explicitly incorporate the feedback-induced shift of the potential. 
We derive the corresponding optimal feedback law 
and analytically obtain the minimum achievable steady-state phonon occupation number. 
We then compare its cooling performance with that of LPF feedback, 
demonstrating that the redesigned LQG control achieves 
a lower phonon occupation number. 
Finally, 
we compare the redesigned and conventional LQG cost functions, 
clarifying how they target distinct physical control objectives.

The rest of this paper is organized as follows. 
In Sec.~\ref{sec:achievable-E}, 
we formulate the redesigned LQG control problem, 
derive the corresponding optimal feedback law, 
and evaluate the minimum achievable steady-state phonon occupation number 
on the basis of the stochastic quantum master equation. 
In Sec.~\ref{sec:achievable-E-others}, 
we briefly summarize the corresponding result for LPF feedback. 
In Sec.~\ref{sec:Comparison}, 
we compare the minimum phonon occupation numbers 
achievable with the redesigned LQG control and the LPF feedback, 
and discuss their dependence on the detection efficiency. 
We then compare the conventional and redesigned LQG cost functions 
in terms of their physical control objectives. 
In Sec.~\ref{sec:conclusion}, 
we summarize our findings. 
In Appendix~\ref{sec:LQR-with-CWC}, 
we review the LQR theory 
for a cross-weighted quadratic cost function, 
which is applied in Appendix~\ref{sec:appendix-optimal-control-law-QHO} 
to derive the optimal feedback law used in Sec.~\ref{sec:achievable-E}.

\section{Achievable Minimum Energy for Redesigned Linear--Quadratic--Gaussian Control} \label{sec:achievable-E}

In this section, 
we consider the optimal feedback cooling of a quantum harmonic oscillator 
by adjusting the minimum of its trapping potential 
on the basis of a continuous position measurement. 
Our primary objective is to minimize the energy of the oscillator 
with respect to the feedback-shifted potential, 
thereby bringing the system as close as possible to 
the ground state of the instantaneous trapping potential.

We first formulate the corresponding optimal control problem 
for a classical harmonic oscillator, 
for which the position and momentum are assumed to be fully accessible 
and measurement backaction is absent. 
We then extend the resulting state-feedback law to the quantum system 
by invoking the separation principle and the certainty equivalence principle. 
Finally, we evaluate the steady-state cooling performance 
in terms of the achievable phonon occupation number.

\subsection{Optimal Feedback Control Scheme for a Classical Harmonic Oscillator}
\label{subsec:opt-cont-cl-harmonic-oscillator}

We first formulate the optimal control problem for a classical harmonic oscillator 
whose trapping-potential minimum is shifted by feedback. 
We assume that both the position and momentum of the oscillator 
are fully accessible without measurement backaction. 
The resulting optimal state-feedback law will be extended 
to the quantum system in the next subsection. 
A detailed derivation is given in Appendix~\ref{sec:appendix-optimal-control-law-QHO}.

We consider a particle of mass $m$ trapped in a one-dimensional harmonic potential with frequency $\omega$. 
Denoting the position and momentum by $x$ and $p$,  
the feedback control implemented by shifting the potential minimum by $X$ is 
described by the Hamiltonian 
\begin{equation}
    H = \frac{p^2}{2m} + \frac{1}{2} m\omega^2 (x-X)^2. 
    \label{eq:classical-Hamiltonian-under-FB}
\end{equation}
The potential shift $X$ is determined on the basis of the particle's position and momentum. 
For the above Hamiltonian \eqref{eq:classical-Hamiltonian-under-FB}, 
the equations of motion of the particle are given by 
\begin{equation}
    \dot x = \frac{p}{m},\quad 
    \dot p = - m\omega^2(x-X).
    \label{eq:EoM-classical-harmonic-oscillator}
\end{equation}

Our primary objective is to minimize the energy of the oscillator 
with respect to the feedback-shifted potential given by 
Eq.~\eqref{eq:classical-Hamiltonian-under-FB}. 
To formulate this physical objective as an optimal control problem, 
we choose the time-integrated energy over a control duration $T$ 
as the cost function:
\begin{equation}
    J_T[X] \coloneqq \int_0^T 
    \left[
        \frac{p^2}{2m} 
        + \frac{1}{2}m\omega^2(x-X)^2
    \right]\d t.
    \label{eq:cost-func}
\end{equation}
Here, the control duration $T$ is taken to be sufficiently long 
compared with the relevant timescales of the oscillator. 
For a finite control duration, 
the optimal feedback law is generally time-dependent. 
However, for a sufficiently long control horizon,  
it approaches 
the stationary solution of the corresponding infinite-horizon LQR problem, 
except near the terminal time. 
Since we are interested in the steady-state cooling performance, 
we focus on this stationary feedback law. 
As derived in Appendix~\ref{sec:appendix-optimal-control-law-QHO}, 
the stationary optimal feedback law is
\begin{equation}
    X=x-\frac{p}{m\omega}.
    \label{eq:optimal-control}
\end{equation}
The optimal control given by Eq.~\eqref{eq:optimal-control} is 
one of the main results of this paper.

\subsection{Minimum Achievable Energy of a Quantum Harmonic Oscillator under the Optimal Control Scheme} \label{subsec:SDEs}

In this subsection, 
we discuss the optimal feedback control for cooling a quantum harmonic oscillator   
and show how the optimal scheme is established 
on the basis of classical control strategy derived 
in Sec.~\ref{subsec:opt-cont-cl-harmonic-oscillator}.

Following the approach of Ref.~\cite{doherty1999feedback}, 
we assume that the position of a quantum harmonic oscillator
is continuously monitored 
with strength $k$ and detection efficiency $\eta \, ( 0 < \eta \leqslant 1)$.
The density operator of the particle conditioned on the measurement outcome obeys 
the stochastic quantum master equation \cite{doherty1999feedback}
\begin{equation}
    \d \hat\rho = -\frac{\i}{\hbar} [\hat H ,\hat \rho]\, \d t 
    + 2k \mathcal{D}[\hat x] \hat \rho \, \d t 
    + \sqrt{2 \eta k} \mathcal{H}[\hat x] \hat \rho \,\d W , \label{eq:SME}
\end{equation}
where $\hat H$ is the Hamiltonian 
\begin{alignat}{1}
    \hat H &= \frac{\hat p^2}{2m} + \frac{1}{2}m\omega^2 (\hat x - X)^2, 
\end{alignat} 
$\d W$ is a Wiener increment, 
and $\mathcal{D}$ and $\mathcal{H}$ are superoperators defined as 
\begin{alignat}{1}
    \mathcal{D} [\hat x] \hat \rho &= \hat x\hat \rho \hat x - \frac{1}{2} \{\hat x^2, \hat \rho\}, \\
    \mathcal{H} [\hat x] \hat \rho &= \{ \hat x - \Tr[\hat x \hat \rho] , \hat \rho\}.
\end{alignat}
We shall discuss the dynamics of the particle in terms of expectation values of 
the position and the momentum and their (co)variances:
\begin{gather}
    \langle \hat z \rangle \coloneqq \Tr[\hat z \hat \rho], \quad V_z \coloneqq \Tr[\hat z^2 \hat \rho ] - \langle \hat z \rangle^2, \label{eq:expec_vari} \\
    C \coloneqq \frac{1}{2} \langle \hat x\hat p + \hat p\hat x \rangle - \langle \hat x \rangle \langle \hat p \rangle \label{eq:covari}, 
\end{gather}
where $z$ represents either $x$ or $p$.
For a Gaussian initial state, 
the quadratic Hamiltonian combined with the continuous linear position measurement 
preserve the Gaussian profile of the conditional state~\cite{gardiner2004quantum}. 
Consequently, the conditional means and (co)variances defined in 
Eqs.~\eqref{eq:expec_vari} and \eqref{eq:covari} evolve according to  
the following stochastic and ordinary differential equations \cite{doherty1999feedback}:
\begin{alignat}{1}
    \d \langle \hat x\rangle &= \frac{\langle \hat p\rangle}{m}\d t + 2\sqrt{2 \eta k} V_x \d W, \label{eq:EoM_x}\\
    \d \langle \hat p\rangle &= - m\omega^2 (\langle \hat x \rangle - X) \d t + 2\sqrt{2 \eta k} C \d W, \label{eq:EoM_p} \\
    \frac{\d V_x}{\d t} &= \frac{2C}{m} - 8\eta k (V_x)^2, \label{eq:EoM_Vx}\\
    \frac{\d V_p}{\d t} &= -2m\omega^2 C + 2k \hbar^2 - 8 \eta k C^2, \label{eq:EoM_Vp} \\
    \frac{\d C}{\d t} &= \frac{V_p}{m} - m\omega^2 V_x - 8 \eta k V_x C  \label{eq:EoM_Covari}.
\end{alignat}

To explicitly incorporate the feedback-shifted potential into the optimization, 
we adopt the following cost function:
\begin{alignat}{1}
    J_{\text{LQG},T}^\text{(red)}[X]
    &\coloneqq
    \mathbb{E}
    \left[
        \int_0^T
        \left\langle 
            \frac{\hat p^2}{2m}
            +\frac{1}{2}m\omega^2(\hat x-X)^2
        \right\rangle
        \d t
    \right] \\
    &=
    \mathbb{E}
    \left[
        \int_0^T
        \left(
            \frac{\langle\hat p\rangle^2}{2m}
            +\frac{1}{2}m\omega^2(\langle\hat x\rangle-X)^2
        \right)
        \d t
    \right] \notag \\
    &\qquad + \int_0^T \left(
        \frac{V_p}{2m}
        +\frac{1}{2}m\omega^2V_x
    \right) \d t.
    \label{eq:cost-func-rev-LQG}
\end{alignat}
Here, $\mathbb{E}$ denotes the ensemble average over measurement records. 
In deriving the second equality, 
we have taken the covariance terms outside the ensemble average, 
using the fact that $V_x$ and $V_p$ are independent of the measurement record 
[Eqs.~\eqref{eq:EoM_Vx}--\eqref{eq:EoM_Covari}]. 
Furthermore, since Eqs.~\eqref{eq:EoM_Vx}--\eqref{eq:EoM_Covari} 
do not depend on the feedback displacement $X$, 
the contribution of the covariances to the cost function 
is independent of the control $X$ 
and thus does not affect the optimal feedback law.

The dynamics of the conditional means 
[Eqs.~\eqref{eq:EoM_x} and \eqref{eq:EoM_p}] 
form a linear stochastic system driven by additive Gaussian noise, 
while the control-dependent part of the cost function is quadratic 
in both the conditional means and the control input. 
Therefore, the separation principle and the certainty equivalence principle 
apply to the present LQG problem~\cite{wiseman2009quantum}. 
Consequently, the optimal feedback law can be obtained 
from the corresponding deterministic control problem 
by replacing $x$ and $p$ with their conditional means.
For a sufficiently long control duration, 
the feedback law approaches the stationary solution, yielding 
\begin{equation}
    X
    =
    \langle\hat x\rangle
    -\frac{\langle\hat p\rangle}{m\omega}.
    \label{eq:Q_opt-control_minimun-point}
\end{equation}
While the conditional means in 
Eq.~\eqref{eq:Q_opt-control_minimun-point} 
are not directly obtainable from the measurement record, 
they can be estimated in real time using the quantum Kalman filter, 
as discussed in Sec.~\ref{subsec:Avail-SysPara}.

To evaluate the steady-state cooling performance under  
Eq.~\eqref{eq:Q_opt-control_minimun-point}, 
we take the long-time limit where the conditional covariances 
reach their stationary values. 
Setting the left-hand sides of 
Eqs.~\eqref{eq:EoM_Vx}--\eqref{eq:EoM_Covari} to zero yields~\cite{doherty1999feedback}
\begin{alignat}{1}
    C &= \frac{\hbar}{2\sqrt{\eta}} \sqrt{\frac{\xi-1}{\xi+1}}, \label{eq:steady-state-value-C} \\
    V_x &= \frac{\hbar}{\sqrt{2\eta} m\omega} \frac{1}{\sqrt{\xi+1}}, \label{eq:steady-state-value-Vx} \\
    V_p &= \frac{m\hbar \omega}{\sqrt{2\eta}} \frac{\xi}{\sqrt{\xi+1}}, \label{eq:steady-state-value-Vp}
\end{alignat}
where 
\begin{equation}
    \xi := \sqrt{1 + \eta \tilde \gamma^2},\quad 
    \tilde \gamma := \frac{4 \hbar k}{m \omega^2}.
\end{equation}
Under the feedback control scheme given by 
Eq.~\eqref{eq:Q_opt-control_minimun-point}, 
the dependence of the energy on the conditional means 
reduces to that on $\langle\hat p\rangle$ alone: 
\begin{alignat}{1}
    E =
    \left\langle
        \frac{\hat p^2}{2m} + \frac{1}{2}m\omega^2 (\hat x-X)^2
    \right\rangle  = 
    \frac{V_p}{2m} + \frac{1}{2}m\omega^2 V_x + \frac{\langle \hat p \rangle^2}{m}. \label{eq:E-expression-optcont}
\end{alignat}
The steady-state energy achieved under the feedback law in Eq.~\eqref{eq:Q_opt-control_minimun-point}
is obtained by taking the long-time limit 
of the ensemble average of Eq.~\eqref{eq:E-expression-optcont}. 
It then remains to evaluate 
$\mathbb{E}[\langle\hat p\rangle^2]$. 
Using It\^o's rule, $\d W^2=\d t$~\cite{gardiner2009stochastic}, 
we obtain
\begin{equation}
    \d\mathbb{E}
    \left[\langle\hat p\rangle^2\right]
    =
    \left(
        -2\omega\mathbb{E}
        \left[\langle\hat p\rangle^2\right]
        +8\eta k C^2
    \right)\d t.
    \label{eq:ODE_E-p2}
\end{equation}
The solution to Eq.~\eqref{eq:ODE_E-p2} gives 
the steady-state value of $\mathbb{E} \left[\langle \hat p \rangle^2\right]$:
\begin{equation}
    \lim_{t\to +\infty} \mathbb{E} \left[\langle \hat p \rangle^2\right] = \frac{4 \eta k C^2}{\omega}.
\end{equation} 
Substituting this result and Eqs.~\eqref{eq:steady-state-value-C}--\eqref{eq:steady-state-value-Vp} 
in Eq.~\eqref{eq:E-expression-optcont}, 
we obtain the energy achievable with the optimal control scheme \eqref{eq:Q_opt-control_minimun-point} as
\begin{equation}
    E=
    \frac{\hbar\omega}{2\sqrt{\eta}} \left[
        \sqrt{\frac{1+\sqrt{1+\eta \tilde \gamma^2}}{2}} 
        + \frac{(\sqrt{1+\eta \tilde \gamma^2}-1)^2}{2\sqrt{\eta }\tilde \gamma}
    \right], \label{eq:achievable-E_redesigned-LQG}
\end{equation}
and the corresponding achievable phonon occupation number is 
\begin{alignat}{1}
    \overline{n} 
    &\coloneqq \frac{E}{\hbar\omega} - \frac{1}{2} \notag \\
    &=
        \frac{1}{2\sqrt{\eta}}\left[
        \sqrt{\frac{1+\sqrt{1+\eta \tilde \gamma^2}}{2}} 
        + \frac{(\sqrt{1+\eta \tilde \gamma^2}-1)^2}{2\sqrt{\eta }\tilde \gamma}
    \right] -\frac{1}{2}
    , \label{eq:achievable-occu_redesigned-LQG} 
\end{alignat}
which gives the minimum steady-state phonon occupation number 
achievable by the feedback law 
for fixed measurement strength $k$ 
and detection efficiency $\eta$.
Equation~\eqref{eq:achievable-occu_redesigned-LQG} constitutes the second main result of this paper.
We note that in the weak-measurement limit 
$\tilde\gamma\searrow0$, 
Eq.~\eqref{eq:achievable-occu_redesigned-LQG} approaches 
the fundamental lower bound 
for feedback cooling based on continuous position measurement 
with detection efficiency $\eta$~\cite{magrini2021real}:
\begin{alignat}{1}
    n_{0} (\eta) &:= \frac{1}{\hbar\omega} 
    \left(\frac{V_p}{2m} + \frac{1}{2}m\omega^2 V_x \right) - \frac{1}{2} \notag \\
    &=\frac{1}{2} \left(\frac{1}{\sqrt{\eta}}-1\right).
\end{alignat}
For fixed $\eta$, 
Eq.~\eqref{eq:achievable-occu_redesigned-LQG} 
can be expanded around $\tilde\gamma=0$ as
\begin{equation}
    n_0
    + \frac{1}{16} \sqrt{\eta} \tilde \gamma^2 
    + \frac{1}{16} \eta \tilde \gamma^3 
    - \frac{5}{256} \eta^{3/2} \tilde \gamma^4 
    + \mathcal{O}(\tilde \gamma^5), \label{eq:achievable-occu_redesigned-LQG-expansion} 
\end{equation}
which will be used in Sec.~\ref{sec:Comparison} for comparison with other methods.

\subsection{Estimation of the Conditional Means with a Quantum Kalman Filter} \label{subsec:Avail-SysPara}

The conditional means 
$\langle\hat x\rangle$ and $\langle\hat p\rangle$ 
appearing in Eq.~\eqref{eq:Q_opt-control_minimun-point} 
are not directly obtainable from the measurement record. 
We assume that the state-preparation procedure specifies 
the initial Gaussian state, and hence its first and second moments. 
Below, we show that these conditional means can be tracked in real time 
from the measurement record via a quantum Kalman filter.

The measurement record $\d Q$ is related to 
$\langle\hat x\rangle$ and the Wiener increment $\d W$ by
\begin{equation}
    \d Q
    =
    \langle\hat x\rangle\d t
    +\frac{\d W}{\sqrt{8\eta k}}.
    \label{eq:measurement-outcome}
\end{equation}
The same Wiener increment appears in 
both the conditional dynamics \eqref{eq:EoM_x} and \eqref{eq:EoM_p} 
and the measurement record \eqref{eq:measurement-outcome}. 
Eliminating $\d W$ from Eqs.~\eqref{eq:EoM_x} and \eqref{eq:EoM_p} yields 
\begin{alignat}{1}
    \d\langle\hat x\rangle
    &=
    \frac{\langle\hat p\rangle}{m}\d t
    +8\eta k V_x
    \left(
        \d Q-\langle\hat x\rangle\d t
    \right),
    \label{eq:filter_x}\\
    \d\langle\hat p\rangle
    &=
    -m\omega^2
    \left(
        \langle\hat x\rangle-X
    \right)\d t
    +8\eta k C
    \left(
        \d Q-\langle\hat x\rangle\d t
    \right).
    \label{eq:filter_p}
\end{alignat}
Given that the initial first and second moments are known 
and the Gaussian form of the conditional state is preserved, 
Eqs.~\eqref{eq:EoM_Vx}--\eqref{eq:EoM_Covari} 
determine the covariances at all subsequent times. 
Together with these covariance equations, 
Eqs.~\eqref{eq:filter_x} and \eqref{eq:filter_p} 
recursively determine the conditional means 
from the measurement record $\d Q$.
Equations~\eqref{eq:EoM_Vx}--\eqref{eq:EoM_Covari}, \eqref{eq:filter_x},  
and \eqref{eq:filter_p}
constitute the quantum Kalman filter for the present Gaussian system.

\section{Achievable Minimum Energy for LPF Feedback}
\label{sec:achievable-E-others}

In this section, 
to provide a basis for comparison with the redesigned LQG control in Sec.~\ref{sec:Comparison}, 
we briefly summarize the LPF feedback scheme proposed in Ref.~\cite{sugiura2025quantum}. 
In this scheme, 
the measurement record is processed by an LPF 
to determine the feedback-induced shift of the trapping-potential minimum. 
Its cooling performance is evaluated using the energy 
with respect to the feedback-shifted potential, 
consistently with the objective considered in Sec.~\ref{sec:achievable-E}.

Under the LPF feedback scheme, 
the potential shift $X$ is generated by passing the measurement record 
through a low-pass filter:
\begin{equation}
    X(t)
    =
    s\int_{-\infty}^{t}
    \e^{-s(t-t')}\d Q(t').
    \label{eq:LPF_method-filter}
\end{equation}
Here, $s>0$ is the cutoff angular frequency of the low-pass filter, 
and the lower limit $-\infty$ corresponds to the stationary filtering regime. 
The value of $s$ that minimizes the steady-state energy is given by~\cite{sugiura2025quantum} 
\begin{equation}
    s= \frac{\tilde \gamma}{\displaystyle \sqrt{\frac{1}{\eta}+\frac{\tilde \gamma^2}{2}}} \omega,
    \label{eq:optimal-s}
\end{equation}
and the corresponding minimum steady-state phonon occupation number 
is obtained as~\cite{sugiura2025quantum}
\begin{alignat}{1}
    n_{\rm LPF} &= \frac{1}{\hbar\omega}\mathbb{E} \left[
        \left\langle
            \frac{\hat p^2}{2m} + \frac{1}{2}m\omega^2 (\hat x - X)^2
        \right\rangle
    \right]-\frac{1}{2} \\
    &= \frac{1}{2} \left(\sqrt{\frac{1}{\eta} + \frac{\tilde \gamma^2}{2}}-1\right) \label{eq:occu-LPF} \\
    &= n_0 + \frac{1}{8} \sqrt{\eta} \tilde{\gamma}^2 
    - \frac{1}{64} \eta^{3/2} \tilde \gamma^4 + \mathcal{O} ( \tilde \gamma^6).\label{eq:occu-LPF-expansion}
\end{alignat}

\section{Discussion} \label{sec:Comparison}

\subsection{Comparison between Redesigned LQG Control and LPF Feedback}

We compare the minimum steady-state phonon occupation numbers 
achievable under the redesigned LQG control 
with those obtained via the LPF feedback.

\begin{figure}[tb]
    \centering
    \includegraphics[width=0.45\textwidth]{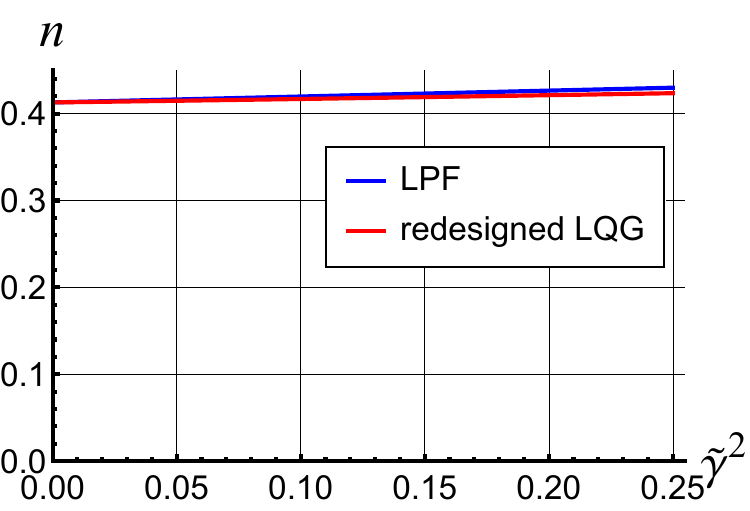}
    \caption{\label{fig:Comparison-Energy_2} 
    Achievable phonon occupation number $n$ as a function of $\tilde\gamma^2$ 
    at a detection efficiency of $\eta=0.3$. 
    The results are shown for both the LPF feedback (Eq.~\eqref{eq:occu-LPF}, blue curve) 
    and the redesigned LQG control 
    (Eq.~\eqref{eq:achievable-occu_redesigned-LQG}, red curve). 
    The two curves are nearly indistinguishable on this scale.
    }
\end{figure}

Figure~\ref{fig:Comparison-Energy_2} shows the two occupation numbers 
for a detection efficiency of $\eta=0.3$, 
which is comparable to that reported in Ref.~\cite{magrini2021real}. 
The redesigned LQG control achieves a lower phonon occupation number 
than that achieved by the LPF feedback; 
however, the difference is small relative to the scale of the occupation numbers themselves.
In the weak-measurement regime, 
Eqs.~\eqref{eq:achievable-occu_redesigned-LQG-expansion} 
and \eqref{eq:occu-LPF-expansion} yield
\begin{equation}
    n_{\rm LPF}
    -
    n_{\rm LQG}^{\rm (red)}
    =
    \frac{1}{16}\sqrt{\eta}\tilde\gamma^2
    +\mathcal{O}(\eta\tilde\gamma^3).
\end{equation}
Thus, for a fixed $\tilde\gamma$, 
the absolute difference between the two schemes decreases 
as the detection efficiency is reduced, 
whereas both occupation numbers increase due to the common contribution $n_0(\eta)$. 
Because the LPF feedback requires only simple first-order low-pass filtering 
rather than the real-time state estimation required for LQG control, 
it offers a straightforward alternative that achieves a cooling performance 
close to that of the optimal redesigned LQG control 
at detection efficiencies attainable in current experiments.

\begin{figure}[tb]
    \centering
    \includegraphics[width=0.45\textwidth]{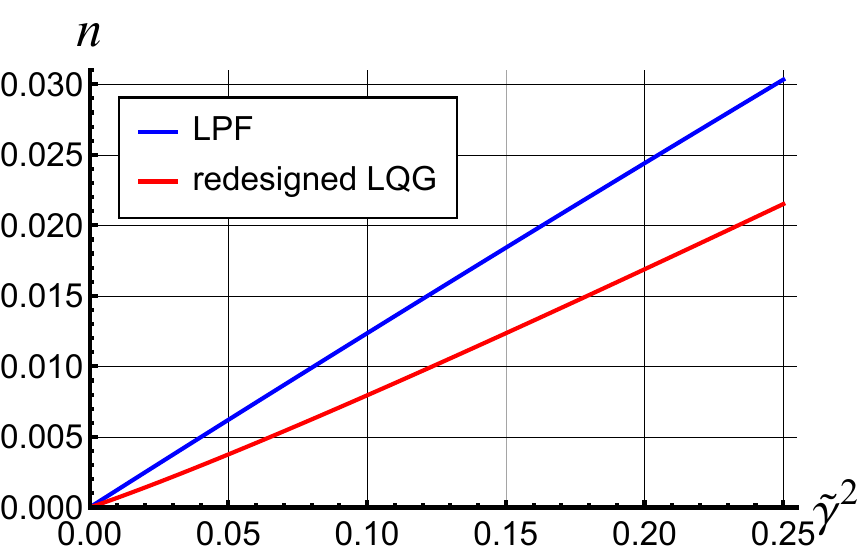}
    \caption{\label{fig:Comparison-Energy_3}
    Achievable phonon occupation number $n$ as a function of $\tilde\gamma^2$ 
    at unit detection efficiency ($\eta=1$), 
    for the LPF feedback (Eq.~\eqref{eq:occu-LPF}, blue curve) 
    and the redesigned LQG control 
    (Eq.~\eqref{eq:achievable-occu_redesigned-LQG}, red curve). 
    The difference between the two schemes is significant  
    compared with that at $\eta=0.3$ shown in Fig.~\ref{fig:Comparison-Energy_2}.
    }
\end{figure}

Figure~\ref{fig:Comparison-Energy_3} shows the corresponding comparison 
under perfect detection ($\eta=1$). 
In contrast to the $\eta=0.3$ case, 
the difference between the two schemes becomes significant. 
Although both occupation numbers approach zero in the formal limit 
$\tilde\gamma\to0$, 
an arbitrarily small measurement strength cannot be realized in practice. 
At experimentally relevant finite measurement strengths, 
the redesigned LQG control therefore provides a significant advantage 
in approaching the motional ground state as the detection efficiency approaches unity---%
a regime that may become accessible in future experiments.

\subsection{Conventional vs Redesigned Cost Functions}

In this subsection, 
we compare the redesigned LQG control with the conventional approach. 
To facilitate a direct comparison with the redesigned cost function introduced in Sec.~\ref{sec:achievable-E}, 
the conventional LQG cost function~\cite{doherty1999feedback} 
is expressed using the notation adopted in this study: 
\begin{alignat}{1}
    J_{\text{LQG},T}^{\text{(con)}}[X]
    &=
    \mathbb{E}
    \left[
        \int_0^T
        \left(
            \left\langle
                \frac{\hat p^2}{2m}
                +\frac{1}{2}m\omega^2\hat x^2
            \right\rangle \right.\right. \notag \\
    &\qquad \qquad \qquad\qquad  \left.\left.
            +\frac{1}{2\Gamma^2}m\omega^2X^2
        \right)
        \d t
    \right].
    \label{eq:conventional-LQG-cost-func}
\end{alignat}
Here, the last term penalizes the feedback displacement $X$, 
with $\Gamma$ determining the relative weight assigned to this control penalty.

In the conventional cost function~\eqref{eq:conventional-LQG-cost-func}, 
the oscillator energy is evaluated with respect to a fixed spatial origin, 
while the feedback displacement $X$ is penalized separately. 
The resulting control therefore favors cooling and localization 
around this fixed reference point while limiting the magnitude of the feedback displacement~\cite{doherty1999feedback}. 
By contrast, 
the potential-energy term in the redesigned cost function is proportional to $(\hat x-X)^2$, 
thereby depending on the particle's displacement 
relative to the instantaneous potential minimum. 
Consequently, the redesigned cost penalizes this relative displacement, 
whereas the conventional cost separately penalizes the particle's displacement 
from the fixed origin and the feedback displacement. 
The two cost functions therefore represent distinct physical control objectives:
cooling and localization with respect to a fixed reference point 
in the conventional LQG control, 
and cooling with respect to the feedback-shifted potential 
in the redesigned LQG control.

The conventional and redesigned controls should therefore not be regarded 
simply as inferior and superior implementations of the same optimization problem. 
The conventional cost is appropriate when localization around a fixed reference point 
and limitations on the feedback displacement are integral to the control objective. 
In contrast, the redesigned cost is suited for situations where the primary objective is 
cooling with respect to the actual feedback-shifted potential. 
In practical experiments, 
additional penalties or constraints may be required. 
For example, 
a penalty on $X$ can account for limitations on the feedback displacement, 
whereas a hard constraint on its magnitude can explicitly represent a finite steering range. 
A weak penalty on the absolute position may also be introduced 
if the particle must remain close to the optical axis.

\section{Conclusion}\label{sec:conclusion}

In this paper, 
we have investigated the feedback cooling of a quantum harmonic oscillator 
where the trapping-potential minimum is shifted by the feedback operation. 
In this setting, 
the oscillator energy relative to the feedback-shifted potential 
is taken as the physical quantity to be minimized. 
To formulate this cooling objective within the LQG framework, 
we have redesigned the cost function to incorporate the feedback-induced shift of the potential. 
Furthermore, we have derived the stationary optimal feedback law given by 
Eq.~\eqref{eq:Q_opt-control_minimun-point} 
and demonstrated that it achieves a lower steady-state phonon occupation number 
than the LPF feedback~\cite{PhysRevE.111.014152,sugiura2025quantum}.

At detection efficiencies attainable in current experiments, 
the difference in the steady-state phonon occupation number 
between the two feedback schemes is minimal 
(Fig.~\ref{fig:Comparison-Energy_2}). 
Because the LPF feedback requires only a low-pass filter 
rather than the real-time state estimation essential for LQG control, 
it offers a simple alternative that achieves cooling performance 
comparable to that of the redesigned LQG control in this regime.
By contrast, 
the difference becomes appreciable as the detection efficiency approaches unity 
(Fig.~\ref{fig:Comparison-Energy_3}). 
Although both schemes approach the ground state 
in the formal limit of vanishing measurement strength, 
an arbitrarily small measurement strength cannot be realized in practice. 
At experimentally relevant finite measurement strengths, 
the redesigned LQG control therefore provides a growing advantage 
in approaching the motional ground state 
as the detection efficiency nears unity.

We have also clarified that 
the conventional and redesigned cost functions represent distinct physical control objectives 
rather than different implementations of the same optimization problem. 
The appropriate cost function must therefore be selected 
according to the intended control objective. 
Crucially, it remains an open question which of these control objectives is more favorable 
for the preparation of nonclassical states, 
such as macroscopic quantum superpositions, entangled states, and Fock states. 
This question depends on the specific requirements of the state-preparation protocol 
and is thus left for future work.

\begin{acknowledgements}
    This work was supported by KAKENHI Grant No.~JP22H01152 from the Japan Society for the Promotion of Science. 
    We gratefully acknowledge the support from the CREST program ``Quantum Frontiers" 
    (Grant No.~JPMJCR23I1) by the Japan Science and Technology Agency.
    K.S.~was supported by KAKENHI Grant No.~JP23KJ0730 from the Japan Society 
    for the Promotion of Science (JSPS) and 
    FoPM, a WINGS Program, the University of Tokyo. 
    M.U.~was supported by the RIKEN TRIP initiative. 
    We thank Kiyotaka Aikawa, Mitsuyoshi Kamba, Masaya Nakagawa, and Akihiro Hokkyo 
    for fruitful discussion.
\end{acknowledgements}

\appendix

\section{LQR Problem for Cross-Weighted Cost} 
\label{sec:LQR-with-CWC}

We consider a linear system whose state is described by 
a column vector $\bm q\in\R^n$ $(n\in\N_{\geqslant 1})$.
The state evolves according to
\begin{equation}
    \dot{\bm q}
    =
    A\bm q+B\bm u,
\end{equation}
where $A\in\R^{n\times n}$ and $B\in\R^{n\times m}$ are constant matrices, 
and $\bm u\in\R^m$ $(m\in\N_{\geqslant 1})$ denotes a control input.

In Sec.~\ref{sec:achievable-E}, 
we consider a finite yet sufficiently long control duration, 
focusing on the stationary feedback law away from the terminal-time region.
To determine this law, 
we analyze the corresponding infinite-horizon LQR problem with the quadratic cost function
\begin{equation}
    J_\infty[\bm u]
    =
    \int_0^\infty
    \left[
        \bm q^\top R_{qq}\bm q
        +
        2\bm q^\top R_{qu}\bm u
        +
        \bm u^\top R_{uu}\bm u
    \right]\d t.
    \label{eq:cost_func-CWCLQR}
\end{equation}
Here, $R_{qq}$, $R_{qu}$, and $R_{uu}$ are matrices 
of sizes $n\times n$, $n\times m$, and $m\times m$.
This formulation is referred to as 
the linear--quadratic regulator (LQR) problem.
When $R_{qu}$ is nonzero, 
the cost function is called \textit{cross-weighted} 
because its integrand contains a cross term 
between the state $\bm q$ and the control input $\bm u$.

We assume the following conditions:
\begin{enumerate}[(i)]
    \item \label{item:avail-state_para} 
    The state $\bm q$ is available for feedback;
    \item 
    $R_{qq}$ is symmetric and positive semidefinite, 
    $R_{uu}$ is symmetric and positive definite, and
    \begin{equation}
        \begin{bmatrix}
            R_{qq} & R_{qu} \\
            R_{qu}^\top & R_{uu}
        \end{bmatrix}
        \succeq0.
    \end{equation}
\end{enumerate}
Here, a symmetric matrix $M\in\R^{\ell\times\ell}$ 
$(\ell\in\N_{\geqslant1})$ 
is positive definite (positive semidefinite) if
\begin{equation}
    \bm v^\top M\bm v>0
    \quad
    \left(
        \bm v^\top M\bm v\geqslant0
    \right)
\end{equation}
holds for every nonzero vector $\bm v\in\R^\ell$, respectively.

For a finite control duration, 
the optimal feedback gain is in general time-dependent 
and determined by a differential Riccati equation~\cite{levine2018control}. 
In the present work, 
we consider a sufficiently long control duration 
to focus on the feedback law away from the terminal-time region. 
In this stationary regime, 
the optimal feedback law is obtained by solving the corresponding 
algebraic Riccati equation~\cite{levine2018control}
and given by 
\begin{equation}
    \bm u = -K\bm q,
\end{equation}
where
\begin{equation}
    K
    =
    R_{uu}^{-1}
    \left(
        R_{qu}^{\top}
        +
        B^{\top}S
    \right).
    \label{eq:LQR-K}
\end{equation}
Here, $S$ is a symmetric positive-semidefinite matrix 
satisfying the algebraic Riccati equation
\begin{equation}
    SA_r
    +
    A_r^{\top}S
    +
    \left(
        R_{qq}
        -
        R_{qu}R_{uu}^{-1}R_{qu}^{\top}
    \right)
    -
    SBR_{uu}^{-1}B^{\top}S
    =
    0,
    \label{eq:ARE_LQR}
\end{equation}
where
\begin{equation}
    A_r
    =
    A
    -
    BR_{uu}^{-1}R_{qu}^{\top}.
\end{equation}
For the system considered in Appendix~\ref{sec:appendix-optimal-control-law-QHO},
we explicitly solve Eq.~\eqref{eq:ARE_LQR}
for the positive-semidefinite stationary solution relevant to the present problem.

\section{Optimal Control Scheme for Cooling a Harmonic Oscillator}
\label{sec:appendix-optimal-control-law-QHO}

In this Appendix, 
we apply the stationary LQR formulation summarized in 
Appendix~\ref{sec:LQR-with-CWC} 
to the feedback cooling of a harmonic oscillator.
As discussed in Sec.~\ref{sec:achievable-E}, 
the control duration is assumed to be sufficiently long, 
and we focus on the stationary feedback law away from the terminal-time region.
We therefore consider the corresponding infinite-horizon optimization problem.

We consider a particle of mass $m$ 
in a harmonic potential with frequency $\omega$.
The energy to be reduced is
\begin{equation}
    H
    =
    \frac{p^2}{2m}
    +
    \frac{1}{2}m\omega^2(x-X)^2,
    \label{eq:total_energy-shift_considered}
\end{equation}
where $x$ and $p$ are the position and momentum of the oscillator, 
and $X$ denotes the feedback-controlled minimum of the potential.
The corresponding infinite-horizon cost function is
\begin{equation}
    J_\infty[X]
    =
    \int_0^\infty
    \left[
        \frac{p^2}{2m}
        +
        \frac{1}{2}m\omega^2(x-X)^2
    \right]\d t.
    \label{eq:cost-func-app}
\end{equation}
Since $X$ is a scalar, 
we introduce the scaled state vector $\bm q$ and control input $u$ defined as
\begin{gather}
    \bm q
    \coloneqq
    \begin{bmatrix}
        \displaystyle \sqrt{\frac{m\omega^2}{2}}\,x &
        \displaystyle \frac{p}{\sqrt{2m}}
    \end{bmatrix}^{\top},
    \qquad
    u
    \coloneqq
    \sqrt{\frac{m\omega^2}{2}}\,X,
    \label{eq:Sys-paras}
    \\
    R_{qq}
    \coloneqq
    \mathbb{I},
    \qquad
    R_{qu}
    \coloneqq
    \begin{bmatrix}
        -1\\
        0
    \end{bmatrix},
    \qquad
    R_{uu}
    \coloneqq
    [1],
    \label{eq:matrices-R-HO}
\end{gather}
where $\mathbb{I}$ is the $2\times2$ identity matrix.
With these definitions, 
the integrand in Eq.~\eqref{eq:cost-func-app} takes the following cross-weighted quadratic form:
\begin{equation}
    \bm q^\top R_{qq}\bm q
    +
    2\bm q^\top R_{qu}u
    +
    u^2
    =
    \frac{p^2}{2m}
    +
    \frac{1}{2}m\omega^2(x-X)^2.
\end{equation}

The equations of motion
\begin{alignat}{1}
    \dot x &= \frac{p}{m},
    \qquad
    \dot p=-m\omega^2(x-X)
\end{alignat}
can be written as
\begin{equation}
    \dot{\bm q}
    =
    A\bm q+Bu,
\end{equation}
where
\begin{equation}
    A
    =
    \begin{bmatrix}
        0 & \omega\\
        -\omega & 0
    \end{bmatrix},
    \qquad
    B
    =
    \begin{bmatrix}
        0\\
        \omega
    \end{bmatrix}.
    \label{eq:matrices-EoM-HO}
\end{equation}
The normalization factors in Eq.~\eqref{eq:Sys-paras} 
are chosen such that the weighting and dynamical matrices 
take the simple forms shown above.
For these matrices given above, 
the algebraic Riccati equation~\eqref{eq:ARE_LQR} reduces to
\begin{alignat}{1}
    0
    &=
    -\omega^2S_{12}^2,
    \label{eq:Riccati-eq-HO1}
    \\
    0
    &=
    \omega S_{11}
    -
    \omega^2S_{12}S_{22},
    \\
    0
    &=
    2\omega S_{12}
    +
    1
    -
    \omega^2S_{22}^2.
    \label{eq:Riccati-eq-HO3}
\end{alignat}
Here, $S_{ij}$ $(i,j=1,2)$ denotes the $(i,j)$ component of $S$.
Equations~\eqref{eq:Riccati-eq-HO1}--\eqref{eq:Riccati-eq-HO3}
give $S_{12}=S_{11}=0$ and $S_{22}=\pm1/\omega$.
Requiring $S$ to be positive semidefinite restricts the solution to $S_{22}=1/\omega$, 
thereby yielding the unique positive-semidefinite solution:
\begin{equation}
    S
    =
    \begin{bmatrix}
        0 & 0\\
        0 & 1/\omega
    \end{bmatrix}.
    \label{eq:ARE-solution}
\end{equation}
The corresponding feedback-gain matrix obtained from 
Eq.~\eqref{eq:LQR-K} is
\begin{equation}
    K
    =
    \begin{bmatrix}
        -1 & 1
    \end{bmatrix}.
\end{equation}
The stationary feedback law is therefore
\begin{equation}
    u=-K\bm q.
\end{equation}
Rewriting this relation in terms of $x$, $p$, and $X$, 
we finally obtain
\begin{equation}
    X
    =
    x-\frac{p}{m\omega}.
    \label{eq:opt-control_minimun-point-app}
\end{equation}

\bibliography{}

\end{document}